\documentclass[12pt,journal]{article}
\usepackage{graphicx,amssymb,amsmath,amsthm,cite,cases}
\usepackage{color,multirow}
\usepackage{mathrsfs}
\usepackage{setspace}
\usepackage{adjustbox}
\usepackage{array}

\def\til{\tilde}
\def\oj{\hat{\operatorname{J}}}

\def\Dc{\D^{\rm{C}}}

\def\Dw{\D^{\rm{W}}}

\def\vf{\mathbf{f}}
\def\vfr{\mathbf{f}^{\rm{r}}}
\def\vfaq{\mathbf{f}_q^{\kq}}
\def\vfq{\mathbf{f}_q}
\def\vd{\mathbf{d}}

\def\vr{\mathbf{r}}
\def\vW{\mathbf{w}}
\def\vWt{\til{\mathbf{w}}}

\def\vB{\mathbf{b}}
\def\vs{\mathbf{s}}
\def\vst{\mathbf{\til{s}}}
\def\st{\til{s}}
\def\ellt{\til{\ell}}

\def\vc{\mathbf{c}}
\def\vct{\mathbf{\til{c}}}

\def\ctq{\til{c}^\Delta}

\def\Sti{st_{\rm{ind}}}
\def\Stc{st_{\rm{cf}}}
\def\Sts{st_{\rm{sg}}}

\def\Di{\mathbf{D}}
\def\Ba{\mathbf{B}}

\def\Hf{\mathbf{Hf}}
\def\CWC{\mathbf{CW4}}
\def\CWL{\mathbf{CW2}}
\def\CDF{\mathbf{CDF97}}

\def\CDFF{\mathbf{CDF53}}

\def\Db{\mathbf{Db4}}
\def\Coif{\mathbf{Coif}}
\def\Sym{\mathbf{Sym}}
\def\Shq{\mathbf{Short3}}

\def\Nq{N_b}
\newcommand{\la}{\langle}
\newcommand{\ra}{\rangle}

\newcommand{\Span}{{\mbox{\rm{span}}}}

\def\kq{k_q}

\def\be{\begin{equation}}
\def\ee{\end{equation}}
\def\ben{\begin{eqnarray}}
\def\een{\end{eqnarray}}
\def\R{\mathbb{R}}
\def\Z{\mathbb{Z}}

\def\V{V}
\def\W{W}
\def\ov{\overline}

\def\SR{\mathrm{SR}}
\def\sr{\mathrm{sr}}

\def\PRD{\mathrm{PRD}}
\def\CR{\mathrm{CR}}
\def\QS{\mathrm{QS}}
\def\PRDN{\mathrm{PRDN}}

\def\std{\mathrm{std}}
\def\prd{{\mathrm{prd}}}
\def\prdo{{\mathrm{prd}_0}}

\def\std{\mathrm{std}}

\def\D{{\cal D}}

\def\dicv{{\cal V}}
\def\dicw{{\cal W}}

\title{Wavelet Based Dictionaries for Dimensionality 
Reduction of ECG Signals}
\author{Laura Rebollo-Neira\\
Mathematics Department
Aston University\\
B3 7ET, Birmingham, UK\\
\vspace{0.1cm}\\
Dana \v{C}ern{\'a} \\
Technical University of Liberec\\
Liberec, Czech Republic} 
\begin{document}
\maketitle
\baselineskip = 1.5\baselineskip
\begin{abstract}
Dimensionality reduction of ECG signals is considered 
within the framework of sparse representation.
The approach constructs the signal model by selecting 
elementary components from a redundant dictionary via 
a greedy strategy.
The proposed wavelet dictionaries are built 
from the multiresolution scheme, but 
 translating the prototypes within a shorter step 
than that corresponding to the wavelet basis. 
The reduced representation of the signal is shown to be 
suitable for compression at low level distortion. 
 In that regard, compression results 
are superior to previously reported benchmarks on the 
MIT-BIH Arrhythmia data set.
\end{abstract}

\section{Introduction}
Electrocardiography (ECG) is the process of
recording the electrical activity of the heart over
a period of time. The overall goal of the
test is to obtain information about the structure
and function of that organ. Dimensionality 
reduction of ECG signals is relevant to 
techniques for analysis 
and classification of this type of data. Such techniques 
  are subject of ongoing research \cite{BM14, AAT16, 
SM16, LSCC16, TTF16, GC16, LYW16, AQC18,BUG18},   
 which includes methodologies based on 
 the theory of
compressive sensing \cite{MKA11,ZJM13,PCBV15,PBR16,PP18}.

Compressive sensing enhances the concept of sparsity 
by associating it to a 
new framework for digitalization 
\cite{Bar07,CW08}. 
Producing sparse representation of ECG signals 
is the central aim of the present work. 
This implies to represent an ECG 
record as superposition of as few elementary 
components as possible. The same goal was traditionally
 accomplished by disregarding the least significant
 terms in the  wavelet transform of the signal \cite{MD14}.
 However, as shown in this work, much less 
elementary components are necessary if they are 
selected from a redundant
wavelet dictionary rather than from a wavelet basis. 
This allows us to project an ECG record onto a 
subspace of lower dimensionality, in comparison to the 
dimension of the signal, and still reconstruct the 
record at low level distortion. 

The main contributions of the paper are listed below:

\begin{itemize}
\item A number of wavelets dictionaries, based on
existing families of wavelet bases, are developed.

\item Each dictionary is shown to render much higher
levels of sparsity for representing ECG signals
than the corresponding basis. The
tests are performed on the MIT-BIH Arrhythmia
data set consisting of 48 records each of which of 30 min
length.

\item The metric of local sparsity is shown to be
relevant to the detection of 
 non-stationary noise or significant distortion in 
the patters of an ECG record.

\item A strategy for storing the reduced representation of
an ECG signal is considered. The 
resulting file is shown to render higher compression
 ratio than previously reported results within the state of 
the art for the same data set.

\item MATLAB software for constructing the dictionaries, 
as well as for reproducing results, 
has been made available on a dedicated website 
\cite{webpage}.

\end{itemize}
The paper is organized as follows: Sec.~\ref{SM}
presents all the elements to build the proposed 
model for piecewise dimensionality reduction of an 
ECG record.
 Sec.~\ref{ES} describes a strategy 
to encode the reduced representation of the signal.
Sec.~\ref{NT} presents and  discusses numerical results.
The conclusions are summarized in Sec.~\ref{Con}. 

\section{Piecewise ECG signal model} 
\label{SM}
Assuming that an ECG record is given as an array
 $\vf$ of dimension $N$, we divide the
 signal into $Q$ disjoint segments
 $\vfq \in \R^{\Nq}, \,q=1,\ldots, Q$
 and construct an approximation $\vfaq$ for each segment
 as the $\kq$-term `atomic decomposition'
\be
\label{wmodel}
\vfaq=\sum_{n=1}^{\kq} c_q(n)\vd_{\ell^q_n},
\ee
where the elements $\vd_{\ell^q_n}$, called `atoms', are
chosen from a redundant dictionary $\D=\{\vd_n, \in \R^{\Nq}, \|\vd_n\|=1\}$ through a greedy pursuit
strategy. Here for the selection process we adopt the
Optimized Orthogonal Matching Pursuit (OOMP) method
\cite{RNL02,LRN16} which is stepwise optimal in the sense of
minimizing the residual norm $\|\vf_q -\vfaq\|$ at
each iteration. The algorithm  evolves
by selecting the atoms one by one as 
described next.
\subsection{The OOMP Method}
Since ECG records are normally superimposed to  
a smooth background, we initialize the OOMP
algorithm by assuming that the constant atom $\vd_1$ is 
always one of the elements in \eqref{wmodel}. 
Hence, for each $q$ we set $\kq=0$, 
$\Gamma_q=\emptyset$, $\ell_{1}^q=1$, and
$\vW_1^q=\vB_1^{1,q}=\vd_1$.  Accordingly  
$\vf_q^1= \vd_1 \la \vf_q , \vd_1 \ra$ (where $\la \cdot, \cdot \ra$ indicates the Euclidean inner product), 
 and $\vr_q^1= \vf_q - \vf_q^1$. 

From the above initialization the OOMP approach for 
selecting from $\D$  the
elements  $\vd_{\ell^{q}_n},\,n=2,\ldots,\kq$ in
\eqref{wmodel}, and calculating the corresponding
coefficients $c_q(n),\,\,n=1,\ldots,\kq$,
iterates as follows.
\begin{itemize}
\item[1)] Upgrade the set $\Gamma_{q} \leftarrow  \Gamma_{q} \cup 
\ell_{\kq+1}^q$, increase $\kq\leftarrow \kq +1$, and
 select the index of a new atom for the approximation as
\be
\label{oomp}
\ell_{\kq+1}^{q}=\operatorname*{arg\,max}_{\substack{n=1,\ldots,M\\ n\notin \Gamma_{q}}}
 \frac{|\la \vd_n,\vr_{q}^{\kq}
\ra|^2}{1 - \sum_{i=1}^{\kq}
|\la \vd_n ,\vWt_i^{q}\ra|^2},
\ee
with $\vWt_i^{q}= \frac{\vW_i^{q}}{\|\vW_i^{q}\|}$.
\item[2)]{\em{Orthogonalization step.}}
Compute the corresponding new vector $\vW_{\kq+1}^{q}$ as
\be
\begin{split}
\label{GS}
\vW_{\kq+1}^{q}= \vd_{\ell_{\kq+1}^{q}} - \sum_{n=1}^{\kq} \frac{\vW_n^q}
{\|\vW_n^q\|^2} \la \vW_n^q, \vd_{\ell_{\kq+1}^{q}}\ra
\end{split}
\ee
including, for numerical accuracy,  the
re-orthogonalization step:
\be
\label{RGS}
\vW_{\kq+1}^{q} \leftarrow \vW_{\kq+1}^{q}- \sum_{n=1}^{\kq} \frac{\vW_{n}^q}{\|\vW_n^q\|^2}
\la \vW_{n}^q , \vW_{\kq+1}^{q}\ra.
\ee
\item[3)]
{\em{Biorthogonalization step.}} For $n=1,\ldots,\kq$ upgrade vectors $\vB_n^{\kq,q}$ as
\be
\begin{split}
\label{BW}
\vB_{n}^{{\kq}+1,q}&= \vB_{n}^{{\kq},q} - \vB_{\kq+1}^{{\kq}+1,q}\la \vd_{\ell_{{\kq}+1}^{q}}, \vB_{n}^{\kq+1,q}\ra,\\
\vB_{\kq+1}^{\kq+1,q}&= \frac{\vW_{\kq+1}^q}{\| \vW_{\kq+1}^q\|^2}.
\end{split}
\ee
\item[4)]
Calculate
\ben
\vr_{q}^{\kq+1} &=& \vr_{q}^{\kq} - \la \vW_{\kq+1}^{q}, \vf_{q} \ra  \frac{\vW_{\kq+1}^{q}}{\| \vW_{\kq+1}^{q}\|^2}.
\een
\item[5)] If for a given
 $\rho$ the condition
 $\|\vr_{q}^{\kq+1}\| < \rho$ has been met 
 compute the coefficients
$c_q(n) = \la \vB_n^{\kq+1}, \vf_q \ra,\, n=1,\ldots, \kq+1$
to write the approximation $\vfaq$ of $\vf_q$ as in 
\eqref{wmodel}. Otherwise repeat steps 1) - 5). 
\end{itemize}
Once all the segments $\vf_q$ have been approximated,
the approximation of the whole signal
is obtained by assembling the approximation of
the segments as
$\vfr=\oj_{q=1}^Q \vfaq$, where $\oj$ is the
concatenation operation.

The complexity of the method,   dominated
by the selection of atoms, is  O($K \Nq M$) with
$K=\sum_{q=1}^Q \kq.$

A suitable model for dimensionality reduction of the signal 
 must involve significantly fewer 
atoms in their atomic decomposition than the 
dimension of the signal. The success in achieving 
such a goal depends on the dictionary being used. 
Next we discuss  wavelet dictionaries 
which are fit for the purpose.


\subsection{Wavelet Based Dictionaries}
\label{WD}
The proposed wavelet dictionaries are inspired on 
the construction 
of dictionaries for Cardinal Spline Spaces 
(CSS) as discussed in previous works 
\cite{ARN05, ARN08, RNX10}. 
Let us denote $\V_j, j\geq 0$ to the CSS 
 of order $m$ with simple knots at 
the equidistant partition of the interval $[c,d]$ 
 having distance $2^{-j}$ between two
adjacent knots. A basis for $\V_j$ arises 
from the restriction of the functions 
\be
\label{sca}
  \phi_{j,k}(x):= 2^{j/2} \phi(2^j x-k), \, k \in \Z
\ee
to the interval $[c,d]$, which we expressed as 
 $2^{j/2}\phi(2^j x-k)|_{[c,d]}$.
  The function
$\phi(x)\equiv \phi_{0,0}(x)$ is called scaling function. 
For a CSS $\phi(x)$  
 is the cardinal B-spline of order $m$
associated with the uniform simple knot 
sequence $0,1,\dots,m$ \cite{Sch86} 
\be
  \phi(x)=\frac{1}{m!}\sum_{i=0}^m(-1)^i\binom{m}{i}(x-i)^{m-1}_+,
\ee
where $(x-i)^{m-1}_+$ is equal to $(x-i)^{m-1}$ if $x-i>0$ and 0 otherwise.

The complementary wavelet subspace $\W_j$ on $[c,d]$ is 
 constructed in order to
fulfil
\be
  \V_{j+1}=\V_j\oplus \W_j,\ j\in\Z^+, 
\ee
where $\oplus$ indicates the direct sum, i.e. 
$\V_j \cap \W_j=\{0\}$. 
Hence, 
\be
\V_{j+1}=\V_0 \oplus \W_0 \oplus \W_1\oplus \cdots \W_j.
\ee 
For a given value of $j$ a wavelet  basis for
$\W_j$ arises as
\be\label{psi}
 \psi_{j,k}(x)=2^{j/2}\psi(2^j x-k)|_{[c,d]},\,  k\in \Z
\ee
 and the elimination of some redundancy 
introduced by the cutting process at the borders of the
 interval.  Different constructions of 
 mother wavelets $\psi\equiv \psi_{0,0}$ 
 give rise to different wavelet bases.

Within the multi-resolution framework, dictionaries for 
a CSS are simply constructed from the result asserting that 
if for $l \in \Z^{+}$
\be
\label{dicw}
\dicw_{j,l}=\{2^{j/2}\psi(2^j x-\frac{k}{2^l})|_{[c,d]},\,  k\in \Z\},  
\ee
then \cite{ARN08} 
\be
\Span\{\dicw_{j,l}\}= V_{{j+l}}. 
\ee
This result facilitates a tool for designing 
dictionaries of wavelets of different support 
spanning the same CSS. A multi-resolution-like
dictionary $\D_{j,l}$ spanning $\V_{{j}}$ is 
constructed as
 \be
  \label{dic}
  \D_{j,\ell}=\dicv_{0,l}\cup \dicw_{0,l} \cup \dicw_{1,l} \cup \cdots 
  \cup \dicw_{j-l,l},
\ee
with
\be
\dicv_{0,l}=
\{\phi(x-\frac{k}{2^l})|_{[c,d]},\,  k\in \Z\}.
\ee

\begin{figure}
\begin{center}
\includegraphics[width=8cm]{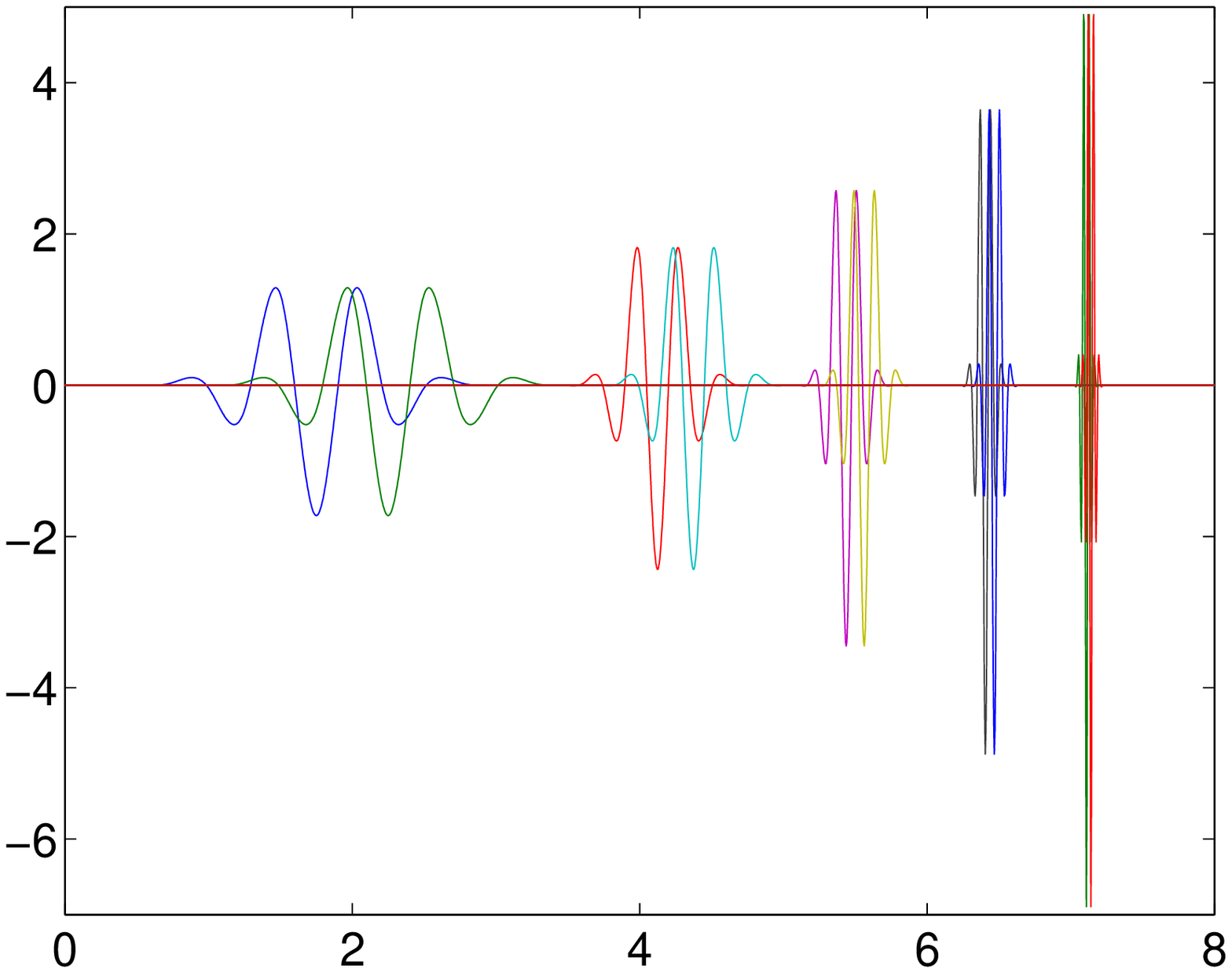}
\includegraphics[width=8cm]{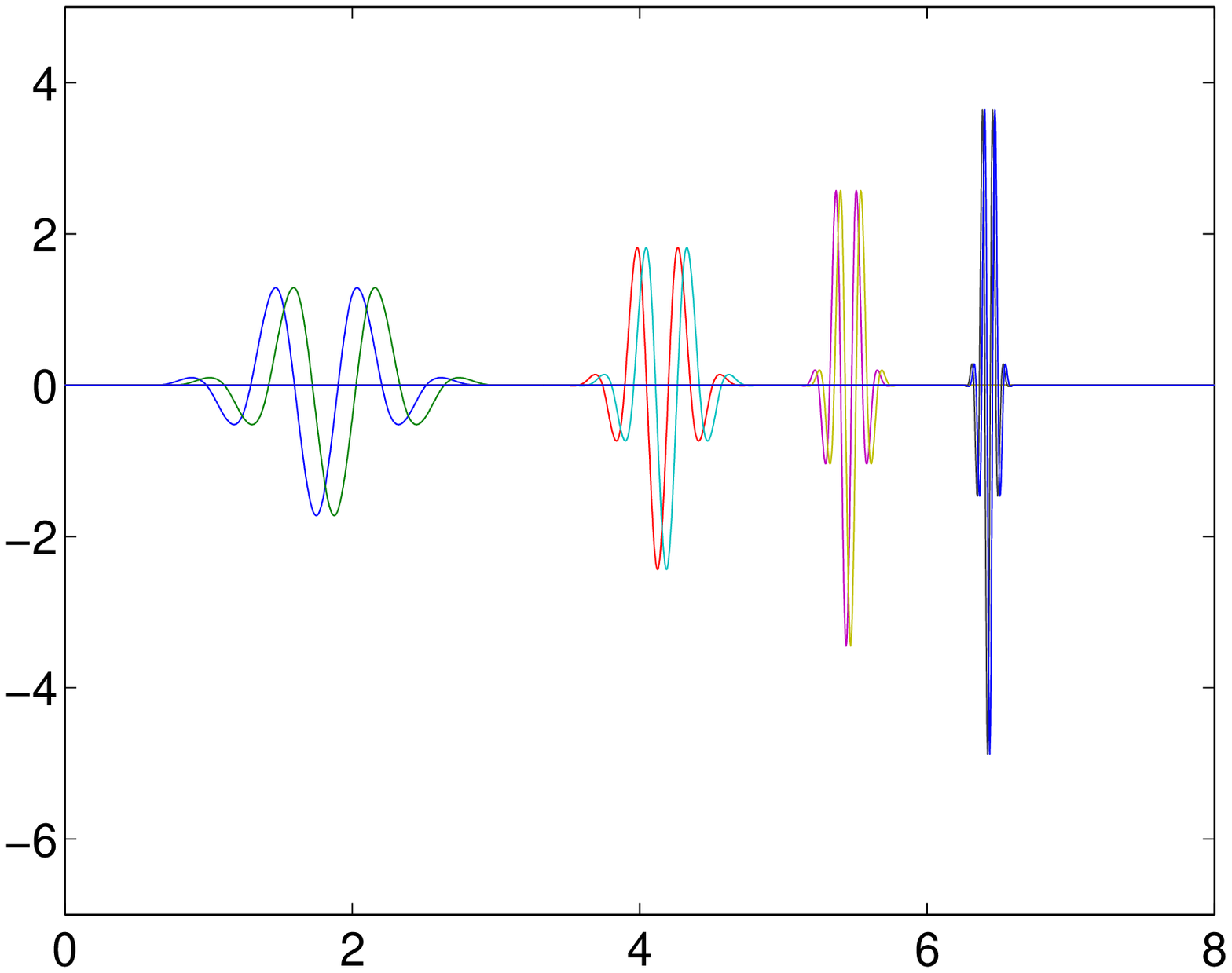}
\caption{Wavelet functions taken from a basis for a CSS 
(left graph)
and from a dictionary spanning the same CSS (right graph).}
\label{Fig2}
\end{center}
\end{figure}

The top graph of Fig.~\ref{Fig2} shows 
consecutive wavelets in a cubic Chui-Wang4 \cite{CW92} 
 wavelet basis 
and consecutive wavelets in a dictionary 
for the same CSS. Notice that 
the wavelet
functions at the finest scale in the dictionary 
 are broader than
those at the finest scale in the basis. 

For the sake of comparison we have built dictionaries for the following wavelet families.

\begin{figure}
\centering
\includegraphics[scale=0.295]{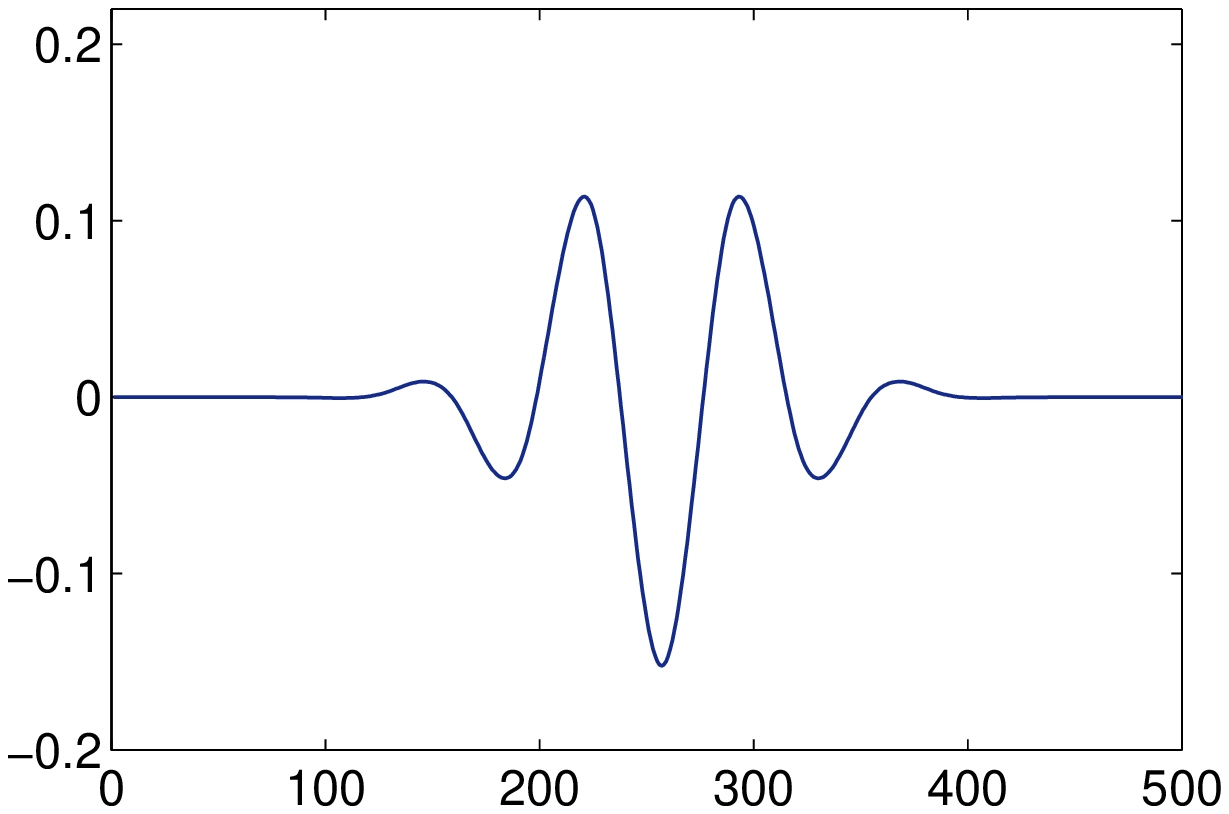} \hspace{-0.2cm}
\includegraphics[scale=0.295]{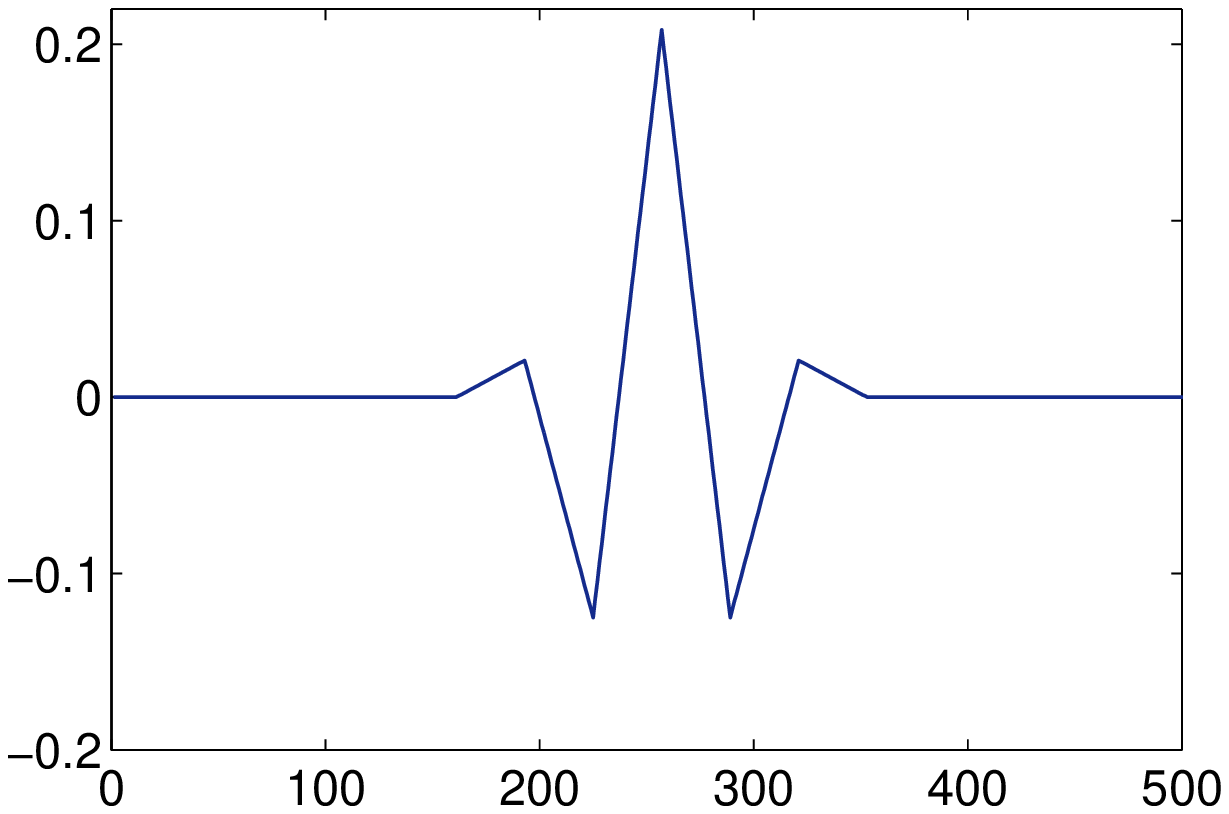}\hspace{-0.2cm}
\includegraphics[scale=0.295]{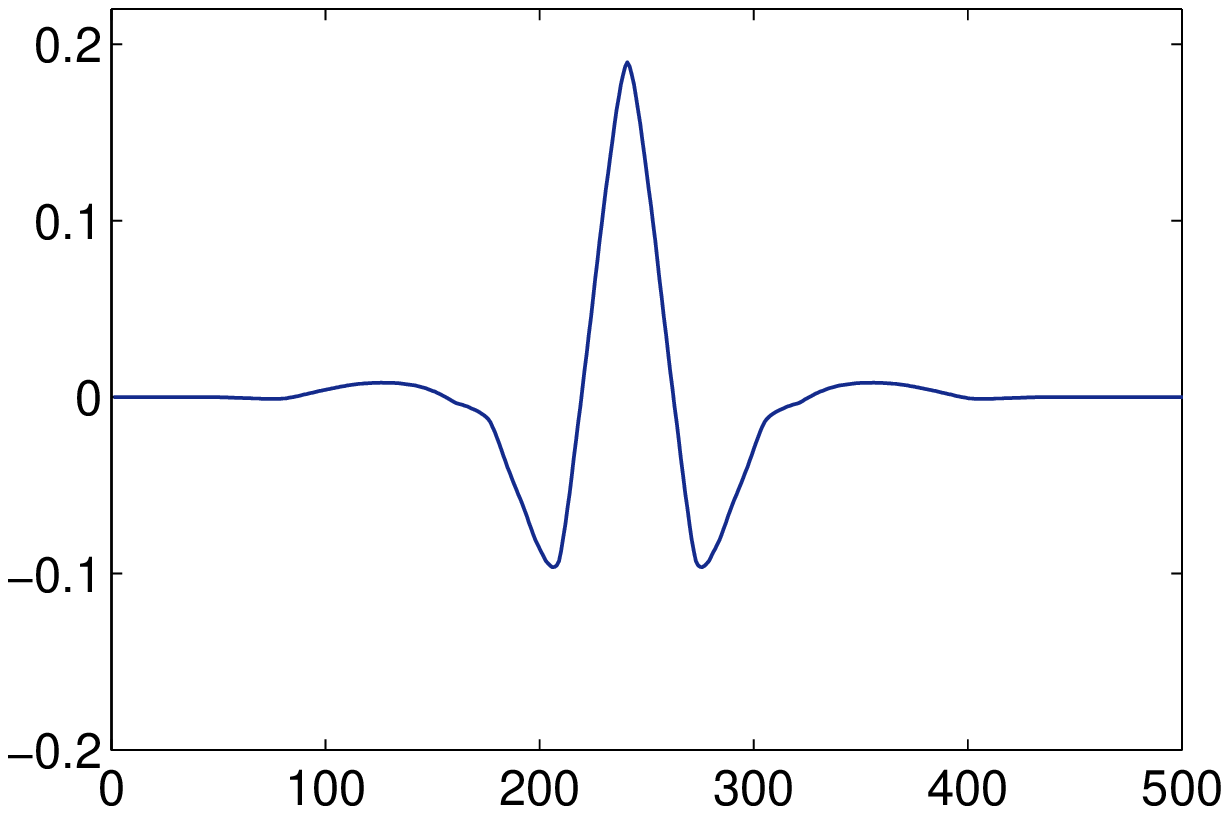} \hspace{-0.2cm}
\includegraphics[scale=0.295]{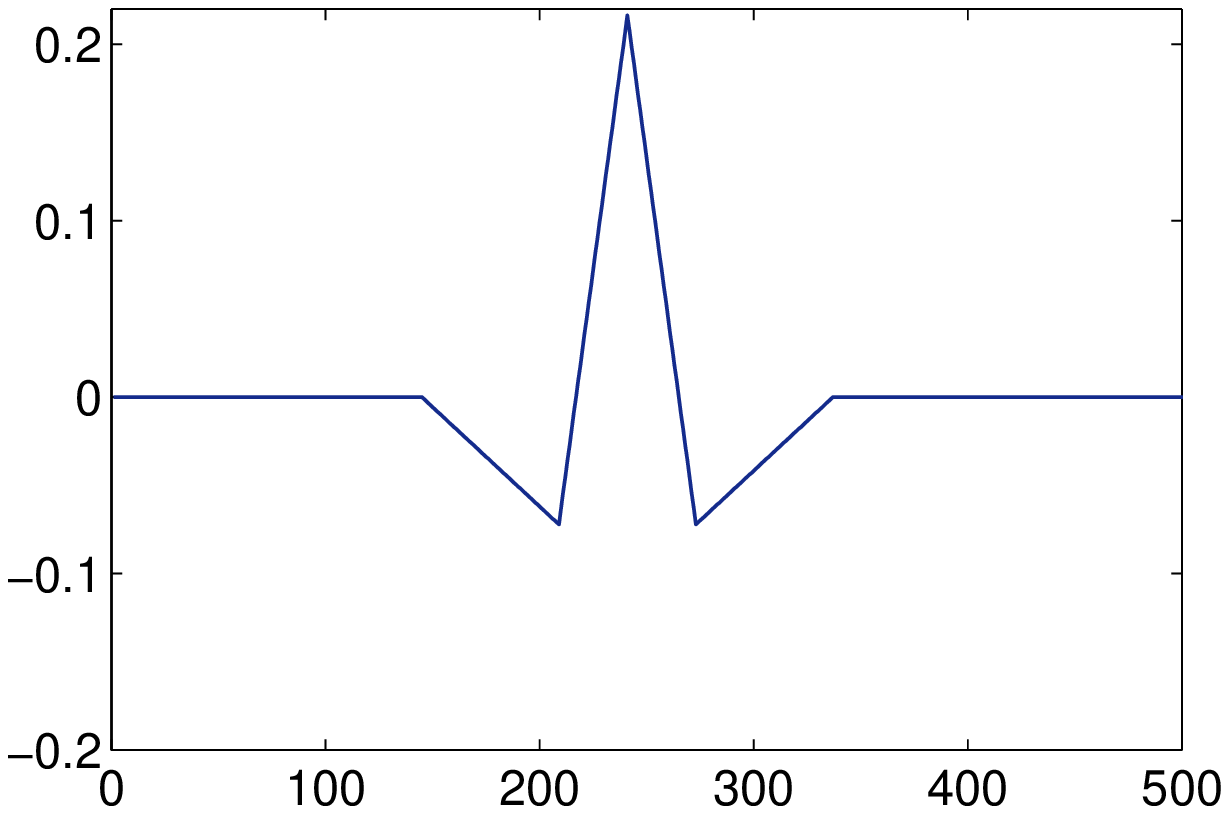}  \hspace{-0.2cm}\\
\includegraphics[scale=0.295]{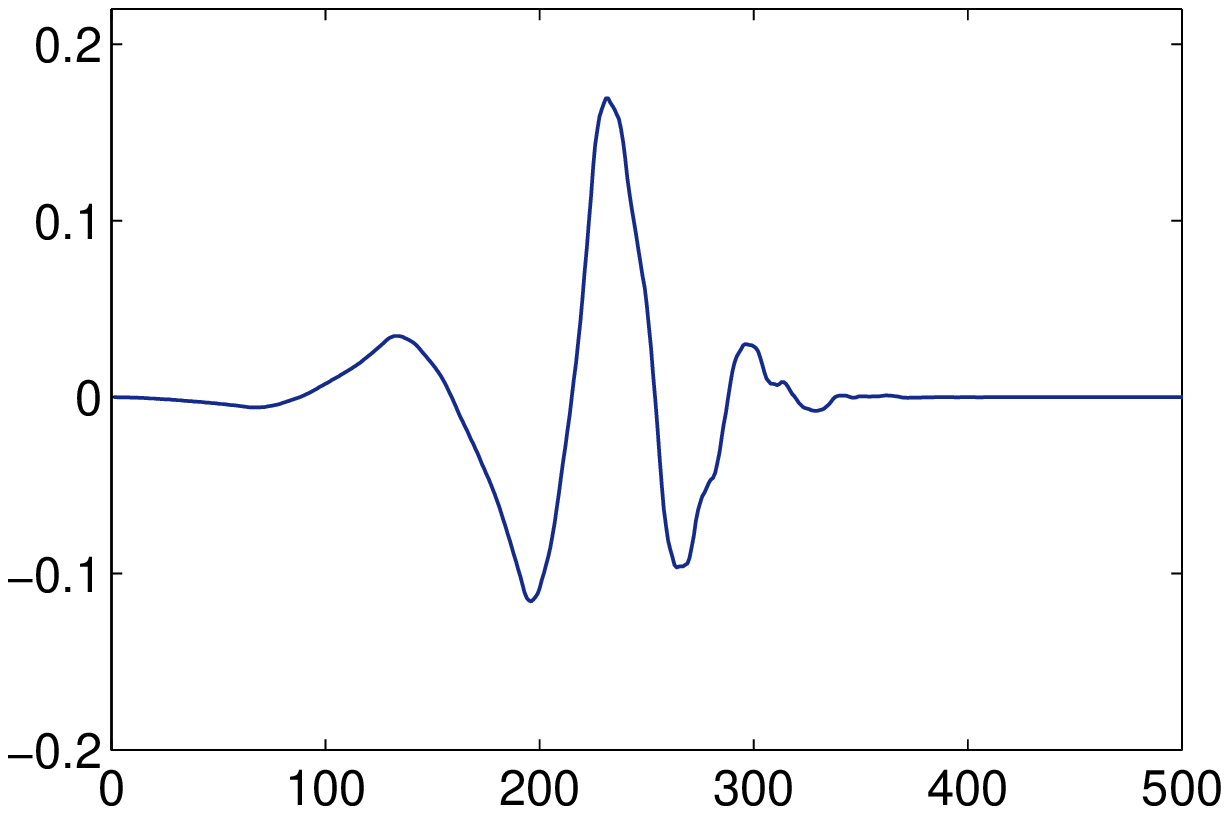} \hspace{-0.2cm}
\includegraphics[scale=0.295]{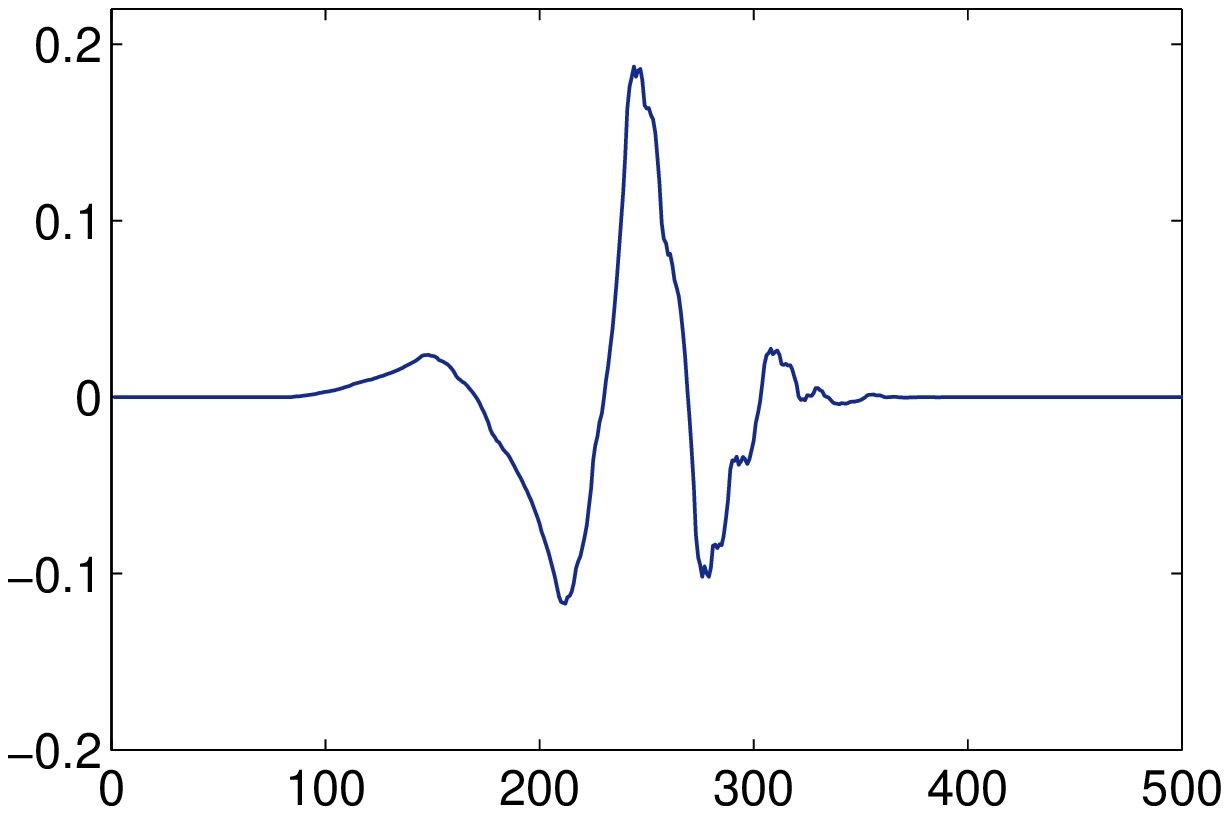}\hspace{-0.2cm}
\includegraphics[scale=0.295]{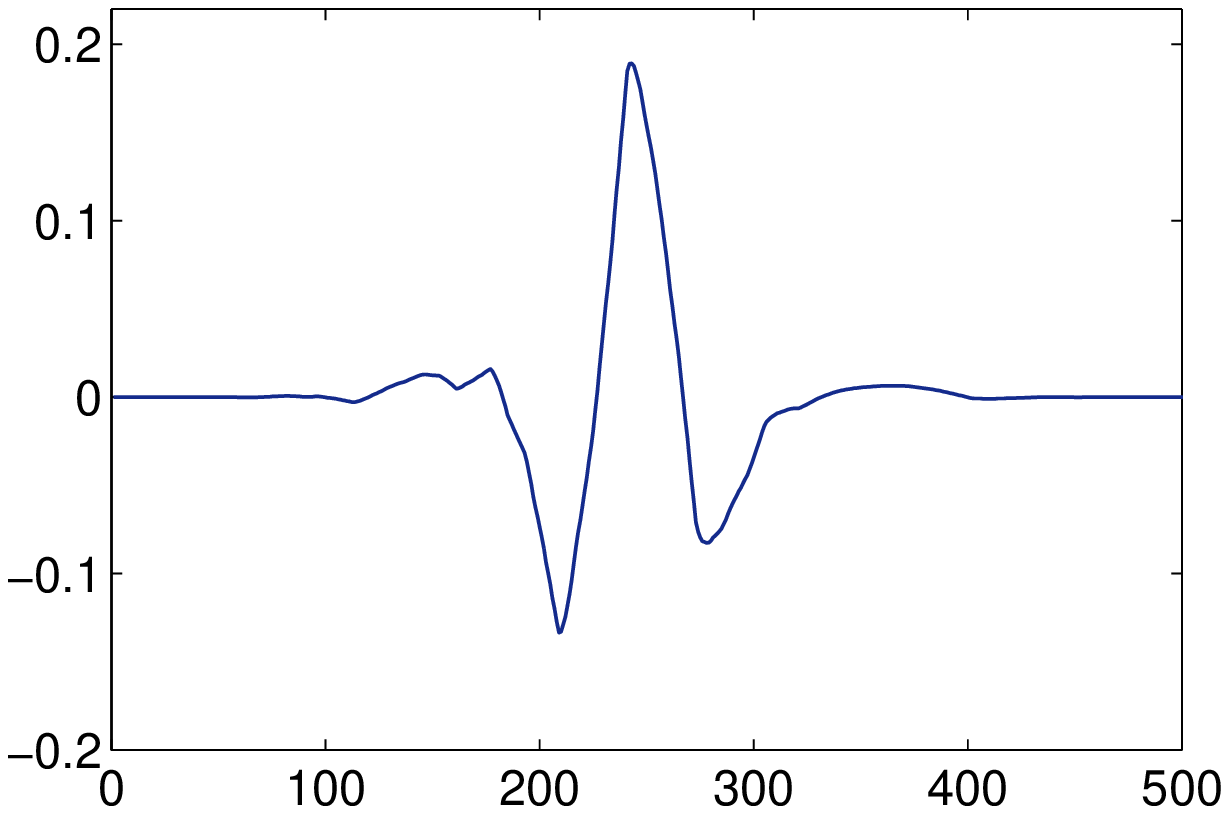}\hspace{-0.2cm}
\includegraphics[scale=0.295]{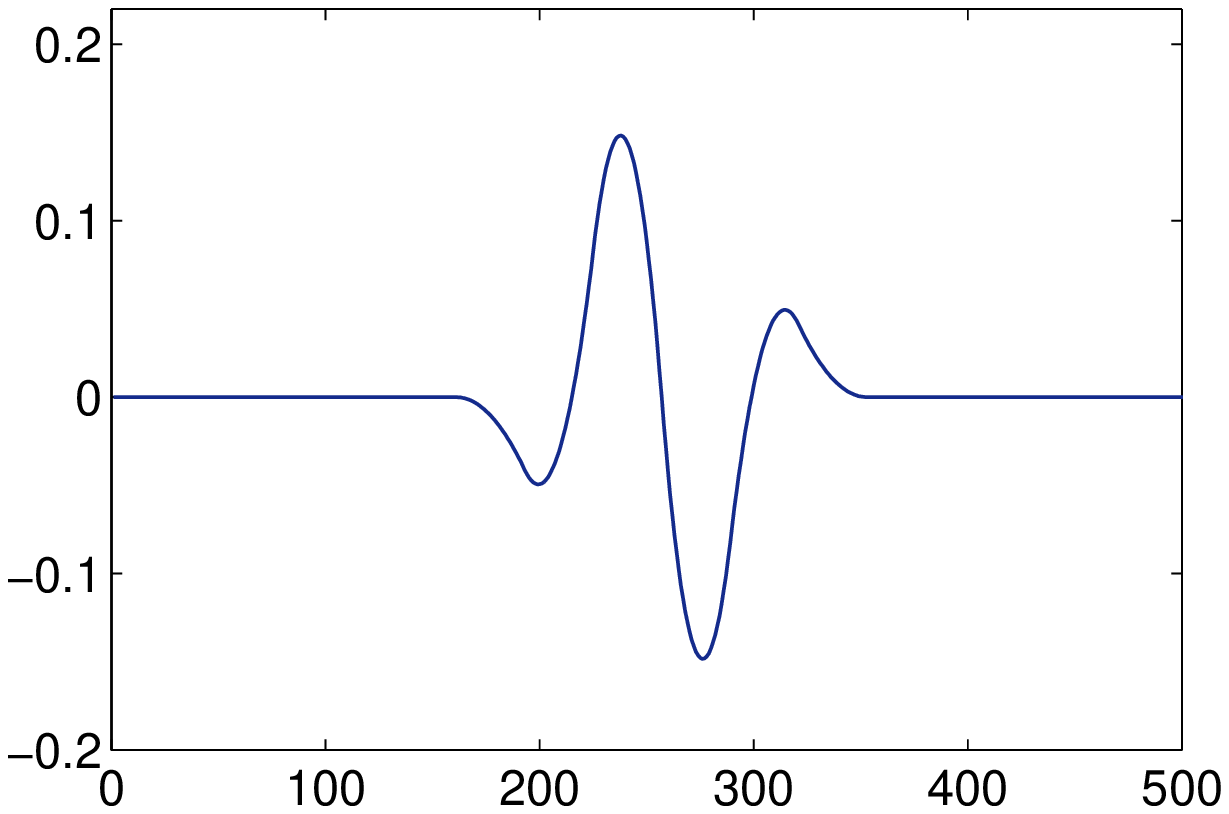}\hspace{-0.2cm}\\
\caption{Wavelet prototypes. From left top to right
bottom the graphs correspond to the families
$\CWC,$ $\CWL,$
$\CDF,$ $\CDFF,$ $\Db,$ $\Coif,$ $\Sym,$
 and  $\Shq$.}
\label{MW}
\end{figure}

{\bf Semi-orthogonal spline wavelets} were mentioned above. Their construction was proposed
by C. Chui and J. Wang in \cite{CW92}. Semi-orthogonal
 wavelets, also called prewavelets, are not orthogonal if the wavelets are at the same scale level, but orthogonality of wavelets at the different levels is preserved. Here we consider the cubic and
linear cases, which are indicated as
 {$\CWC$} and {$\CWL$} respectively.
The wavelet prototypes
corresponding to this family are given the 1st and
2nd  graphs in the 1st row Fig.~\ref{MW}.

{\bf Cohen-Daubechies-Feauveau biorthogonal wavelets} were built in \cite{CDF92}. In this case both primal and dual wavelets have compact support. Wavelets from this family
with 9 and 7 taps filters ({\bf{CDF97}}) and with 5 and 3 taps filters ({\bf{CDF53}}) are used in the JPEG2000
compression standard. The corresponding
 prototypes are  given in the 3rd and 4th graphs in the
 1st  row  of Fig.~\ref{MW}.

{\bf Daubechies wavelets} were built by I. Daubechies as first orthonormal compactly supported continuous wavelets in \cite{Daub88}. They have the shortest possible support for the given number of vanishing moments and they have extremal phase. Here we consider the four vanishing moments and indicate it as {\bf Db4}. This wavelet prototype is displayed in the
 1st
  graph of the 2nd row of Fig.~\ref{MW}.

{\bf Coiflets} were suggested by R. Coifman as orthonormal wavelet systems with
vanishing moments for both scaling functions and wavelets. They were first constructed
by I. Daubechies in \cite{Daub93}. We consider two
vanishing moments used exact values of scaling and wavelet parameters from \cite{CFK08} and
indicate the family as {\bf{Coif}}.
 The mother wavelet is displayed in the
 2nd  graph in the
 2nd  row  of Fig.~\ref{MW}.

{\bf Symlets} are modified version of Daubechies wavelets. They have also the shortest possible support
for the given number of vanishing moments,
see \cite{Daub93}.
We consider four vanishing moments. The prototype wavelet for the family {\bf Sym} is
given in the  3rd  graph in the
 2nd  row  of Fig.~\ref{MW}.

{\bf Spline wavelets with short support} are obtained if the local support of dual wavelets is not required. They were designed in \cite{Chen95, Han06}. Here we consider
quadratic splines and indicate them
as $\Shq$. The prototype for this family is
 plotted in  last graph of the 2nd row of Fig.~\ref{MW}.

In the discrete case dictionaries can be
constructed in a very simple manner:
A redundant dictionary for
the Euclidean space $\R^N$ is simply any set of normalized
vectors $\{ \vd_i \in \R^N,\,\|\vd_i\|=1\}_{i=1}^M$
such that
 $N$ of them are linearly independent and $M>N$.
Inspired by the results for CSS we
construct a dictionary for $\R^N$ simply
by discretization of the functions in the wavelet
dictionary. As already mentioned,  ECG signals are commonly
superimposed to a smooth background, 
hence we enrich a wavelet
dictionary adding a few low frequency
components from a discrete cosine basis. Thus 
the whole dictionary we consider is 
built as $\D=\Dc \cup \Dw$,  with $\Dw$ a discrete wavelet
basis (or dictionary) for $\R^{\Nq}$ and
\ben
\Dc&=&\{w_c(n)
\cos{\frac{{\pi(2i-1)(n-1)}}{2M}},i=1,\ldots,\Nq\}_{n=1}^{M},\nonumber
\een
where each $w_c(n)$ is a normalization constant. 
For the numerical tests we fix $M=10$ for $\Dc$ and for 
$\Dw$ we consider different wavelet dictionaries,
all obtained from the aboved mentioned wavelet bases.

\section{Encoding the signal model}
\label{ES}
For the reduced representation of the signal 
 \eqref{wmodel} 
 to be suitable for storage and transmission purposes
it is necessary to encode it within a small file 
in relation to the size of the original record. 
For this end,  firstly  the 
 absolute value coefficients 
$|c_{q}(n)|,\,n=1\ldots,\kq,\,q=1,\ldots,Q,
$ are
converted into integers as follows:
\be
\label{uniq}
c^\Delta_{q}(n)= \lfloor \frac{|c_{q}(n)|}{\Delta} +\frac{1}{2} \rfloor,
\ee
where $\lfloor x \rfloor$ indicates the largest
integer number
smaller or equal to $x$ and  $\Delta$ is the quantization
parameter.
The signs of the coefficients, represented as
$\vs_{q},\,q=1,\ldots,Q$,
are encoded separately using a binary alphabet.
For creating the strings of numbers to encode the 
 the signal model we proceed as in \cite{RNS17,RN18}.

The indices of the atoms in the
 atomic decompositions of each block $\vf_q$
 are first sorted in ascending
order $\ell_{i}^q \rightarrow \til{\ell}_i^q,\,i=1,\ldots,\kq$,
which guarantees that, for each $q$ value,
$\til{\ell}_i^q < \til{\ell}_{i+1}^q,\,i=1,\ldots,\kq-1$.
The order of the indices induces
order in the unsigned coefficients,
$\vc^\Delta_{q} \rightarrow \vct^\Delta_{q}$ and
in the corresponding signs $\vs_{q} \rightarrow
\vst_{q}$.  The ordered indices are
stored as smaller positive numbers by taking differences
between two consecutive values.
By defining $\delta^q_i=\ellt^q_i- \ellt^q_{i-1},\,i=2,\ldots,\kq$ the follow string stores the indices for block
$q$ with unique recovery
${\ellt^q_1, \delta^q_2, 
\ldots, \delta^q_{\kq}}$.
The number `0' is then used to separate the string
corresponding to different blocks and entropy code
 a long string, $\Sti$, which is built as 
\be
\label{sti}
\Sti=[{\ellt_1}^1,\ldots,\delta^1_{k_1},0,
{\ellt_1}^2,\ldots,\delta^2_{k_2},0, \cdots,
{\ellt_1}^{k_Q},\ldots, \delta^Q_{k_Q}].
\ee
The corresponding quantized magnitude of the coefficients
are concatenated in the strings $\Stc$
 as follows:
\be
\label{stc}
\Stc
= [\ctq_{1}(1), \ldots, \ctq_{1}(k_1), \cdots,
\ctq_{{k_Q}}(1), \ldots, \ctq_{{k_Q}}(k_Q)].
\ee
Using `0' to store a positive sign and `1' to store
negative one, the signs  are placed
in the string, $\Sts$  as
\be
\label{sts}
\Sts=[\st_{1}(1), \ldots, \st_{1}(k_1), \cdots,
\st_{{k_Q}}(1), \ldots, \st_{{k_Q}}(k_Q)].
\ee

\section{Results}
\label{NT}
The numerical tests we present here  use the
 full MIT-BIH Arrhythmia database \cite{MITDB}
 which contains
 48 ECG records. Each of these records is of
30 min length,  consisting of 
 $N=650000$ 11-bit samples at a frequency of 360 Hz.

All our results have been 
 obtained in the MATLAB environment
 using a notebook 2.9GHz dual core i7 3520M CPU and 4GB
of memory.

\subsection{Assessment Metrics}
The quality of the signal approximation is assessed
 with respect to the $\PRD$ calculated as follows,
\be
\PRD=\frac{\|\vf - \vfr\|}{\|\vf\|} \times 100 \%\quad
\ee
where, $\vf$ is the original signal and $\vfr$ is
the signal reconstructed from the approximated segments. 
Since the $\PRD$ strongly depends on the background of 
the signal, the $\PRDN$ as defined below
 is also a relevant metric.
\be
\label{PRDN}
\PRDN=\frac{\|\vf - \vfr\|}{\|\vf - \ov{\vf}\|} \times 100 \%,\quad
\ee
where, $\ov{\vf}$ indicates the mean value of $\vf$.

For a fixed value of $\PRD$ the sparsity of 
a representation is assessed by the sparsity ratio
(SR)
\be
\label{SR}
\text{SR}=\frac{N}{K},
\ee
where $N$ is the total length of the signal 
and $K=\sum_{q=1}^Q \kq,$ with $\kq$ the number of atoms in the
atomic decomposition \eqref{wmodel} of each segment 
of length $\Nq$.
For detection of nonstationary noise in patterns it is useful to 
define the measure of local sparsity $\sr(q)$ as follows:
\be
\label{sr}
\sr(q)=\frac{\Nq}{\kq},\quad q=1,\ldots Q.
\ee
The compression performance depends of the 
size of the file storing the signal representation  
and is assessed by the Compression Ratio (CR) as given by
\be
\text{CR}=\frac{{\text{Size of the uncompressed file}}}{{
\text{Size of the compressed file}}}.
\ee
In the numerical tests below we use the notation $\CR$ to 
denote the compression ratio obtained when 
the string $\Sti, \Stc$ and $\Sts$ are stored  
 in hierarchical data format \cite{HDF5}, 
 simply using the MATLAB instruction
{\tt{save}}. If a 
Huffman coding step is included the corresponding 
compression ratio is denoted as $\CR^{\bf{Hf}}$.

The quality score ($\QS$),
reflecting the tradeoff between
compression performance and reconstruction quality,
is the ratio:
\be
\QS=\frac{\CR}{\PRD}.
\ee
Since the $\PRD$ is a global quantity, in order to
detect possible local changes in the visual quality
of the recovered signal,
 we consider the local $\PRD$ as follows.
Each signal is partitioned in $Q$ segments
$\vf_q,\, q=1\ldots,Q$ of $\Nq$ samples.
 The local $\PRD$ with respect to every segment
in the partition is indicate
as $\prd(q),\,q=1,\ldots Q$ and calculated as
\be
\label{prd}
\prd(q)=\frac{\|\vf_q - \vfr_q\|}{\|\vf_q\|} \times 100 \%,\quad
\ee
where $\vfr_q$ is the recovered portion
of the signal corresponding to the segment $q$.
In all the numerical test of the next section the 
OOMP approximation is stopped through a fixed value 
$\rho$ so as to achieve the 
same value of $\prd$ for all the segments in the 
records.  Assuming that the target $\prd$ before 
quantization is
$\prdo$ we set $\rho = \prdo\|\vf_q\|/100$.

\subsection{Test I}
\label{NT1}
The purpose of the first 
numerical tests is to demonstrate the 
significant gain in dimensionality reduction 
(high values of $\SR$) 
achieved if the component $\Dw$ of the dictionary $\D$ is a 
 wavelet dictionary instead of a wavelet 
basis. The wavelet basis arises using as
 translation parameter $b=1$ and the wavelet 
dictionary using a translation parameter $b=1/4$.
A wavelet basis contains scales corresponding 
to the values $j=3,\ldots,8$ 
while the corresponding dictionary contains one less 
scale i.e. $j=3,\ldots,7$. The exact redundancy of 
the dictionary depends of the wavelet family but 
in all the cases is close to two. The records are 
approximated by partitioning the signal into $Q=1300$ 
segments of $\Nq=500$ data points each.

The second column in Table~\ref{TABLE1} gives the mean 
value $\SR$, after quantization, with respect to the 
48 records in the data set. The standard deviation is 
given in the 3rd column. 
 The approximation 
with all the families is 
accomplished to obtain the same mean value of 
$\PRD$ ($\ov{\PRD}=53$). For this test the 
quantization parameter is fixed. For all the records and 
all the dictionaries $\Delta=35$. 
The 4th column gives the mean value $\CR$ and the 
fifth column the corresponding standard deviation. 
For further improvement in
 $\CR$ we have added an entropy coding step to represent the
strings described in Sec.~\ref{ES} at a fixed bit depth
using  Huffman coding.
The implementation of the step is realized using the off-the-shelf MATLAB function Huff06 available on \cite{Karl}.
 The corresponding outcomes, indicated as $\ov{\CR}^{\Hf}$, 
are given in the sixth column of Table ~\ref{TABLE1}. 
The 1st column lists the different wavelet families which are considered to build the dictionary $\Dw$. The 
subscript B is used to indicate the wavelet basis and 
the subscript D to indicate the 
corresponding dictionary, i.e.  $\CDF_{\Ba}$ indicates the 
 9/7 Cohen-Daubechies-Feauveau biorthogonal wavelet basis
 and $\CDF_{\Di}$ the corresponding wavelet dictionary. 
For further of comparison the last row of the 
table gives the results produced when a 
Fast Wavelet Transform ({\bf{FWT}}) is applied on the 
whole record, as proposed in \cite{LRN19}.  
\begin{table}
\caption{Mean values of  $\SR$ and $\CR$
 with respect to the 48 records in the
  MIT-BIH Arrhythmia database. In all the cases the 
reconstructed signals produce $\ov{\PRD}=53$.}
\label{TABLE1}
\begin{center}
\setlength{\extrarowheight}{0.1cm}
\begin{tabular}{||l||r|r|r|r|r|r|||}
\hline \hline
$\Dw$ &${\ov{\SR}}$ & $\std$ &${\ov{\CR}}$&$\std$ &
$\ov{\CR}^{\Hf}$&$\std$\\ \hline \hline
$\CWC_{\Ba}$& 10.13 & 2.42 & 11.72 & 3.14 &16.70&3.78\\ 
$\CWC_{\Di}$&16.62 & 4.13 & 13.31 & 3.51 &17.80&4.40\\ \hline
$\CWL_{\Ba}$& 11.34 & 2.91 & 12.08 & 3.48&17.78&4.34\\ 
$\CWL_{\Di}$&16.76 & 4.95 & 13.91 & 4.04 &18.37 &5.25\\ \hline
$\CDF_{\Ba}$& 14.59 & 3.90 & 14.90& 4.20 & 20.54 & 4.99\\
$\CDF_{\Di}$&{\bf{20.60}} & 5.79 & {\bf{16.13}}& 4.57 & {\bf{21.26}} & 5.82\\ \hline 
$\CDFF_{\Ba}$& 14.72 & 4.25 & 14.84 & 4.29 & 20.46 & 5.31    \\ 
$\CDFF_{\Di}$&{\bf{20.18}} &6.36 & {\bf{16.02}} & 4.88 & {\bf{21.04}} & 6.27\\ \hline
$\Db_{\Ba}$& 12.81 & 3.28 & 13.49 &3.63 & 18.81 & 4.46\\ 
$\Db_{\Di}$&17.88 & 4.94 & 14.24 &4.05  & 18.97 & 5.15 \\ \hline
$\Coif_{\Ba}$& 11.50 & 2.99 &12.66 &3.47 &17.41 & 4.29 \\ 
$\Coif_{\Di}$&15.27 & 4.57 & 12.58 &3.68 &16.73 & 4.74\\ \hline
$\Sym_{\Ba}$&12.83 & 3.30 & 13.65 &3.68 & 18.96  & 4.47\\
$\Sym_{\Di}$&18.51 & 5.26 & 14.70  &4.22 & 19.26 & 5.77\\ \hline

$\Shq_{\Ba}$&12.78 & 3.18 & 15.09    & 4.31   & 19.68 & 4.59 \\ 
$\Shq_{\Di}$&{\bf{19.84}} & 5.25 & {\bf{16.00}} & 4.24  & {\bf{20.95}} & 5.46 \\ \hline
{\bf{FWT}}\cite{LRN19}&  11.22 & 1.75 & {\bf{23.17}} & 6.57 & {\bf{26.34}} & 6.99 \\ \hline \hline
\end{tabular}
\end{center}
\end{table}

As seen in Table~\ref{TABLE1} the best results in relation 
to dimensionality reduction (largest values of ${\ov{\SR}}$) 
are for $\CDF_{\Di}, \CDFF_{\Di}$ and  $\Shq_{\Di}$ 
and all the wavelet bases
perform poorly in comparison to the dictionaries.
The difference in ${\ov{\CR}}$ is not so pronounced though. 
This is because, in spite of the fact that the approximation 
using a dictionary 
involves much less elementary components in the signal 
decomposition, the dispersion  of  
indices corresponding to the atoms in the decomposition
is larger if using a dictionary. Notice that, as far as 
compression is concerned using a FWT to transform the 
whole signal is more effective than a piecewise 
transformation.  This is true not only 
in terms of the achieved ${\ov{\CR}}$ but also 
in regard to computational time (the approach 
advanced in \cite{LRN19} is very fast). Contrarily, as 
indicated by the values of ${\ov{\SR}}$ in Table~\ref{TABLE1} the method in \cite{LRN19} is not as effective 
for dimensionality reduction as it is for compression. 
However, dimensionality reduction is relevant in its 
own right to methodologies for classification and 
recognition tasks\cite{WSN13,RR18}.

\begin{table}[h!]
\caption{Performance metrics 
for each of the 48 records in the
  MIT-BIH Arrhythmia database
listed in the first column. }
\label{TABLE2}
\begin{center}
\begin{tabular}{||r|r|r|r|r|r||}
\hline \hline
Rec & $\PRD$&$\SR$ &$\CR^{\Hf}$& $\QS$& $\PRDN$
\\ \hline \hline
100 &  0.51& 27.19& 28.27&  55.75&   12.64 \\
101 &  0.51& 25.39& 25.89&  50.82&   9.45 \\
102 &  0.51& 24.43& 24.74&  48.27&   13.13 \\
103 &  0.52& 21.93& 22.06&  42.73&  7.86 \\
104 &  0.53& 19.00& 19.52&  37.21&   10.27 \\
105 &  0.53& 17.37& 18.15&  34.22&    6.43 \\
106 &  0.53& 18.63& 18.87&  35.97&   7.07 \\
107 &  0.55& 12.64& 12.66&  23.01&  3.18 \\
108 &  0.52& 21.13& 22.74&  43.52&   8.51 \\
109 &  0.52& 19.32& 19.42&  37.05&   5.16 \\
111 &  0.51& 23.22& 24.32&  47.45&   9.90 \\
112 &  0.53& 22.72& 24.47&  46.18&   10.20 \\
113 &  0.52& 20.21& 19.80&  38.18&   6.26 \\
114 &  0.50& 32.21& 33.13&  66.17&   14.52 \\
115 &  0.53& 20.18& 20.52&  38.98&   6.73 \\
116 &  0.58& 12.33& 12.90&  22.10&   3.72 \\
117 &  0.52& 27.31& 28.82&  55.27&  9.33 \\
118 &  0.59& 11.06& 12.51&  21.22&  5.90 \\
119 &  0.56& 16.15& 16.50&  29.78&  4.44 \\
121 &  0.51& 33.27& 33.61&  66.10&  7.29 \\
122 &  0.56& 16.42& 17.62&  31.80&  6.50 \\
123 &  0.53& 21.94& 22.93&  43.11&  7.94 \\
124 &  0.53& 23.28& 23.09&  43.92&  4.94 \\
200 &  0.53& 15.56& 16.69&  31.33&  7.05 \\
201 &  0.49& 34.72& 35.05&  71.66&  12.52 \\ 
\hline \hline
\end{tabular}
\begin{tabular}{||r|r|r|r|r|r||}
\hline \hline
Rec & $\PRD$&$\SR$ &$\CR^{\Hf}$& $\QS$& $\PRDN$
\\ \hline \hline
202 &  0.51& 26.24& 26.28&  51.87&   8.44 \\
203 &  0.55& 12.75& 13.86&  25.14&   5.54 \\
205 &  0.51& 25.25& 26.83&  52.54&   12.50 \\
207 &  0.51& 25.93& 25.66&  50.80&   7.08 \\
208 &  0.54& 15.09& 15.69&  29.16&   5.55 \\
209 &  0.54& 15.65& 16.83&  31.40&   9.92 \\
210 &  0.51& 23.84& 24.62&  48.18&   9.72 \\
212 &  0.56& 12.12& 13.56&  24.40&   8.35 \\
213 &  0.56& 11.71& 12.06&  21.71&   4.09 \\
214 &  0.52& 19.22& 19.15&  36.59&   5.54 \\
215 &  0.55& 14.07& 15.34&  28.20&   9.60 \\
217 &  0.53& 16.86& 16.47&  31.06&   4.30 \\
219 &  0.54& 18.06& 18.00&  33.69&   4.43 \\
220 &  0.53& 19.18& 19.65&  37.07&   7.65 \\
221 &  0.52& 22.73& 23.00&  44.69&   8.50 \\
222 &  0.51& 26.52& 27.53&  54.42&   13.53 \\
223 &  0.53& 18.56& 18.70&  35.09&   6.06 \\
228 &  0.52& 18.97& 20.31&  38.91&   7.53 \\
230 &  0.52& 19.12& 19.47&  37.29&   7.30 \\
231 &  0.51& 25.67& 26.20&  51.84&   9.23 \\
232 &  0.50& 29.72& 31.59&  63.83&   14.91 \\
233 &  0.54& 14.26& 14.81&  27.43&   4.96 \\
234 &  0.52& 19.81& 20.50&  39.28&   7.68 \\ \hline
{\bf{mean}}&{\bf{0.53}} & {\bf{20.60}} & {\bf{21.26}}&
{\bf{40.76}} & {\bf{7.99}} \\ \hline
{\bf{std}}& {\bf{0.02}}  &{\bf{5.79}} & {\bf{5.82}} &{\bf{12.47}}& {\bf{2.92}} \\ \hline \hline
\end{tabular}
\end{center}
\end{table}

%
%
%

For further information in Table~\ref{TABLE2}
we provide
the figures of the evaluation metrics for each  of
the 48 records in the dataset. 
 The 2nd column of Table~\ref{TABLE2} shows
the values of $\PRD$ for the records listed in 
the 1st column. The 3rd column shows the values of
$\SR$ after quantization.
 The $\CR^{\Hf}$ is given in the
4th column and the corresponding $\QS$ in the 5th
 column. The last column corresponds to the values
of $\PRDN$.
All the results are obtained
using the dictionary $\CDF_{\Di}$.  The average time
 for producing  Table~\ref{TABLE2} is
 30 s per record.

\begin{figure}[h]
\begin{center}
\includegraphics[width=8.75cm, height=5cm]{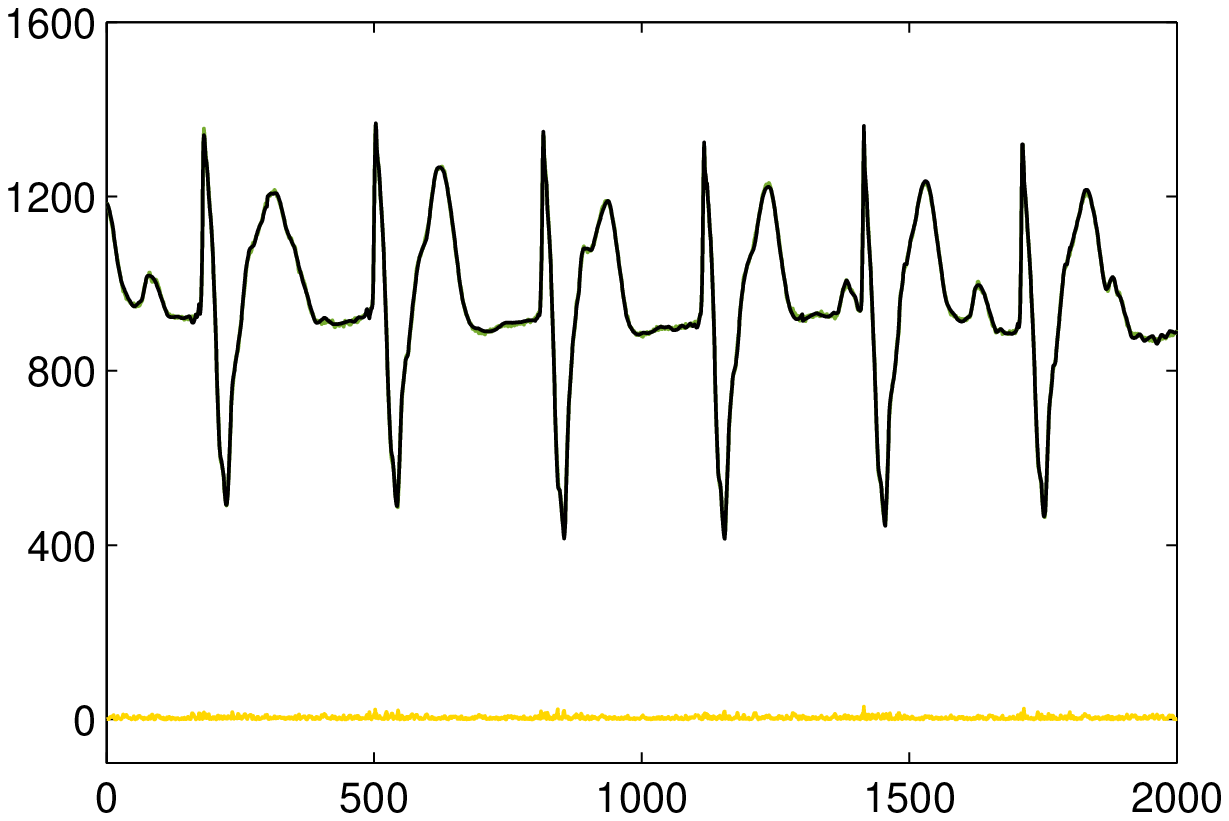}
\includegraphics[width=8.75cm, height=5cm]{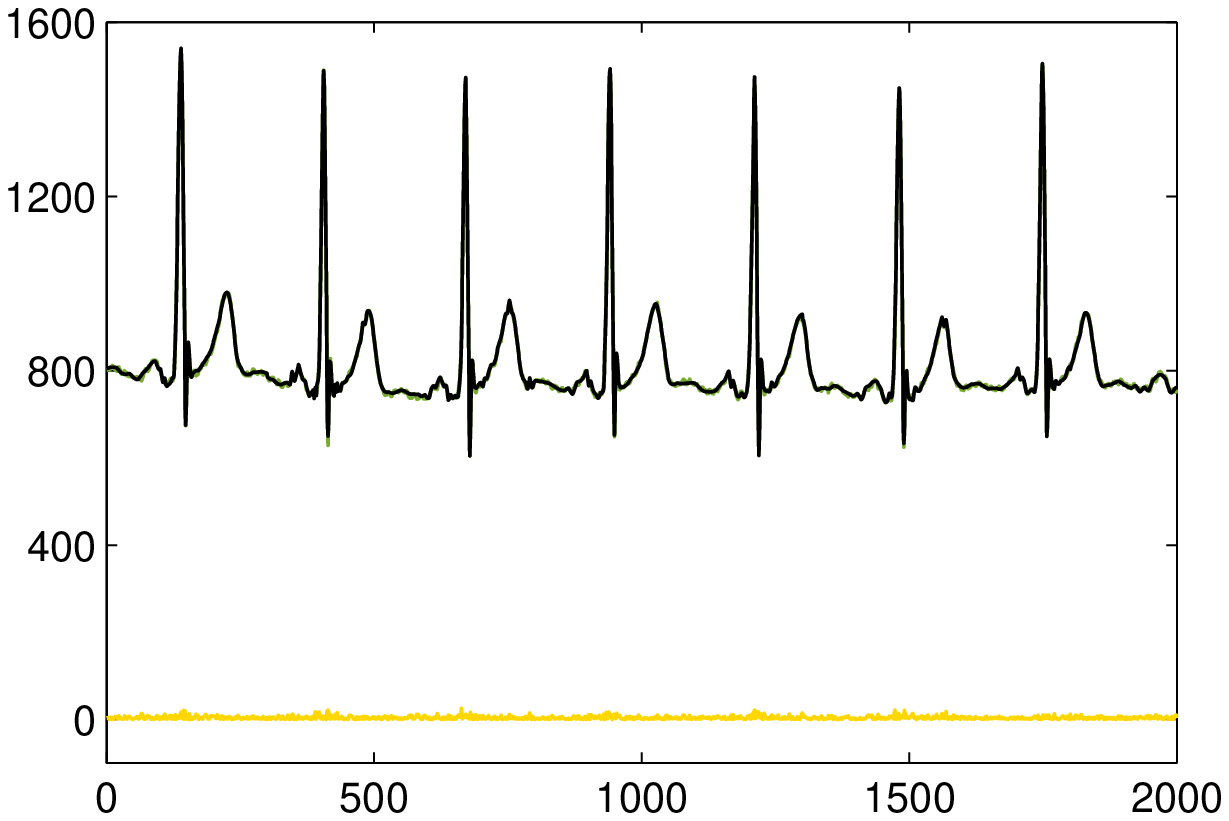}
\end{center}
\caption{The waveforms in the graphs
 are the raw data and the approximations corresponding 
to 2000 points in the records 107 (left) and 116 (right).
The bottom lines  represent the absolute value of
the difference between the data and the approximation.
}
\label{Segs}
\end{figure}

Fig.~\ref{Segs} depicts the approximation and 
raw data (indistinguishable in the scale of the graphs) 
of 2000 points in the records 107 and 116.

\subsection{Test II}
\label{NT2}
In this test we compare compression results with
previously reported benchmarks on the MIT-BIH Arrhythmia
database. 
Since the 
PRD strongly depends on the
baseline in the ECG record, when using the 
relation CR vs PRD
for comparison one needs to be able to 
confirm that the values of
PRD reported in previous work either, do not include
any subtraction of baseline, or the subtracted value
is also reported. From the
information given in the papers we are
comparing with (the relation between
the values of PRD and PRDN) we can guarantee 
that the results correspond to the same database \cite{MITDB}
and without any subtraction of baseline. 
In the range $[0.31 \; 0.67]$ the comparison is 
realized with the classical benchmark \cite{LKL11}.
This 
method is based on peak detection to perform 
the Discrete Cosine Transform (DCT) between 
consecutive peaks and Huffman encoding  for 
lossless compression.
 In the range $[0.80 \;1.71]$ 
the comparison is with the methods in  \cite{MZD15}
and \cite{TZW18}. Both these methods are based on Adaptive Fourier Decomposition (AFD) and 
overperform those in \cite{LKL11}  and \cite{PSS16} for 
values of $\PRD$ in the range being considered.
In \cite{TZW18} the lossless compression 
is realized using Huffman coding while 
 \cite{MZD15} applies symbol substitution. 
Comparisons are also produced with the results arising 
when a FWT is applied on the full record \cite{LRN19}.

Table~\ref{TABLE3} compares CRs obtain with  
 $\CDF_{\Di}$ and 
$\CDFF_{\Di}$ 
for values of $\PRD$ in the range $[0.31 \; 0.81]$, 
Table~\ref{TABLE4} for values in the range $[0.91 \; 1.06]$, 
and Table~\ref{TABLE5} in the range $[0.14 \; 1.71]$. 
The last two rows in all the tables give the values of the
quantization parameter, $\Delta$, and the approximation 
parameter $\prdo$. 
These parameters are set in order to reproduce 
 the same values of  
$\PRD$ as in the publications we are comparing with. 
\begin{table}[h!]
\caption{Comparison between the average compression
performance of the proposed method and the methods
in \cite{LKL11},  
\cite{MZD15}, \cite{TZW18} and \cite{LRN19}, 
for values
of ${\ov{\PRD}}$ in the range [0.31 \; 0,81].}\label{TABLE3}
\begin{center}
\setlength{\extrarowheight}{0.15cm}
\begin{tabular}{|l||r|r|r|r|r||}
\hline \hline
${\ov{\PRD}}$&0.81&0.80& 0.67& 0.48& 0.31\\
\hline \hline
${\ov{\CR}}$\cite{LKL11}&  & & 11.30 & 9.28 & 6.22\\
${\ov{\CR}}$\cite{MZD15}& &18.00 &  & &   \\ 
${\ov{\CR}}$\cite{TZW18}&16.85& &  & &   \\ \hline
${\ov{\CR}}$\cite{LRN19}&33.32&33.05& 27.79& 19.84&10.59 \\ 
$\ov{\CR}^{\Hf}$\cite{LRN19}&36.54&36.26&31.01&23.09&15.00\\ \hline
${\ov{\CR}}\,\,\,\,\,\,\CDF_{\Di}$ &24.03&23.81&20.52 &14.79& 9.37\\ 
$\ov{\CR}^{\Hf}\, \CDF_{\Di}$ &31.64& 31.07& 26.71& 19.59&12.50 \\ \hline
${\ov{\CR}}\,\,\,\,\,\,\CDFF_{\Di}$ &24.76& 24.50 &20.67& 14.67 &8.89 \\ 
$\ov{\CR}^{\Hf}\, \CDFF_{\Di}$&32.59& 31.93& 26.01&19.25&11.93 \\ \hline \hline
$\prdo$& 0.750   &0.740  &0.645 & 0.435 & 0.275\\
$\Delta$&60 & 60 & 40 & 30 & 16 \\ \hline \hline
\end{tabular}
\end{center}
\end{table}


\begin{table}[h!]
\caption{Comparison between the average compression
performance of the proposed method and the methods in 
\cite{MZD15}, \cite{TZW18}, \cite{PSS16} and \cite{LRN19}, 
for values 
of ${\ov{\PRD}}$ in the range $[0.91 \; 1.06]$.}\label{TABLE4}
\begin{center}
\setlength{\extrarowheight}{0.15cm}
\begin{tabular}{|l||r|r|r|r|r||}
\hline \hline
${\ov{\PRD}}$&1.06&1.05& 1.03& 0.94& 0.91\\
\hline \hline
${\ov{\CR}}$\cite{MZD15}& &25.667&  &  &22.27\\ 
${\ov{\CR}}$\cite{TZW18}& &  &22.80 &20.38&  \\
${\ov{\CR}}$\cite{PSS16}&18.59&  &  & &   \\ \hline
${\ov{\CR}}$\cite{LRN19}&41.31&41.24&41.00& 38.52 &37.68\\ 
$\ov{\CR}^{\Hf}$\cite{LRN19}&44.42& 44.23& 44.04& 41.93 &41.08\\ \hline
${\ov{\CR}}\,\,\,\,\,\,\CDF_{\Di}$&30.36&30.12&29.44 &27.29& 26.27\\ 
$\ov{\CR}^{\Hf}\, \CDF_{\Di}$&40.25&39.96&39.11&35.94&34.42\\ \hline
${\ov{\CR}}\,\,\,\,\,\,\CDFF_{\Di}$&31.93& 31.66& 30.84&28.46 &27.28 \\ 
$\ov{\CR}^{\Hf}\, \CDFF_{\Di}$&42.73& 42.39& 41.40&37.81&35.70 \\ \hline \hline
$\Delta$& 100 & 100 & 100 & 80 & 80 \\ 
$\prdo$& 0.940& 0.930 & 0.900&0.850 &0.805\\ \hline \hline
\end{tabular}
\end{center}
\end{table}


\begin{table}[h!]
\caption{Comparison between the average compression
performance of the proposed method and the methods
in 
\cite{MZD15}, \cite{TZW18}, and \cite{LRN19} for values 
of ${\ov{\PRD}}$ in the range $[1.14 \; 1.71]$.}\label{TABLE5}
\begin{center}
\setlength{\extrarowheight}{0.15cm}
\begin{tabular}{|l||r|r|r|r|r||}
\hline \hline
${\ov{\PRD}}$&1.71&1.47&1.29& 1.18& 1.14\\
\hline \hline
${\ov{\CR}}$\cite{MZD15}&38.46&33.85&  &28.21 &  \\
${\ov{\CR}}$\cite{TZW18}&42.27&3.53&30.21& &25.99 \\\hline
${\ov{\CR}}$\cite{LRN19}&62.48&56.78&49.60&47.04&45.75 \\    
$\ov{\CR}^{\Hf}$\cite{LRN19}&63.92& 58.59  & 52.19 & 49.59&48.47 \\ \hline
${\ov{\CR}}\,\,\,\,\,\,\CDF_{\Di}$&52.01&43.83&37.63 & 33.88& 32.69\\
$ \ov{\CR}^{\Hf}\, \CDF_{\Di}$&66.99& 56.44& 48.77& 44.45&42.98\\ \hline
${\ov{\CR}}\,\,\,\,\,\,\CDFF_{\Di}$&57.19& 47.43& 40.66 & 36.27 & 34.62 \\
$\ov{\CR}^{\Hf}\, \CDFF_{\Di}$&73.27&61.00& 52.56&47.19&45.51\\ \hline \hline
$\Delta$&150& 120 & 110 & 110 & 110 \\ 
$\prdo$&1.660&1.430& 1.220& 1.070& 1.020\\ \hline \hline
\end{tabular}
\end{center}
\end{table}


  


\subsection{Discussion}
\begin{figure}
\begin{center}
\includegraphics[width=5cm]{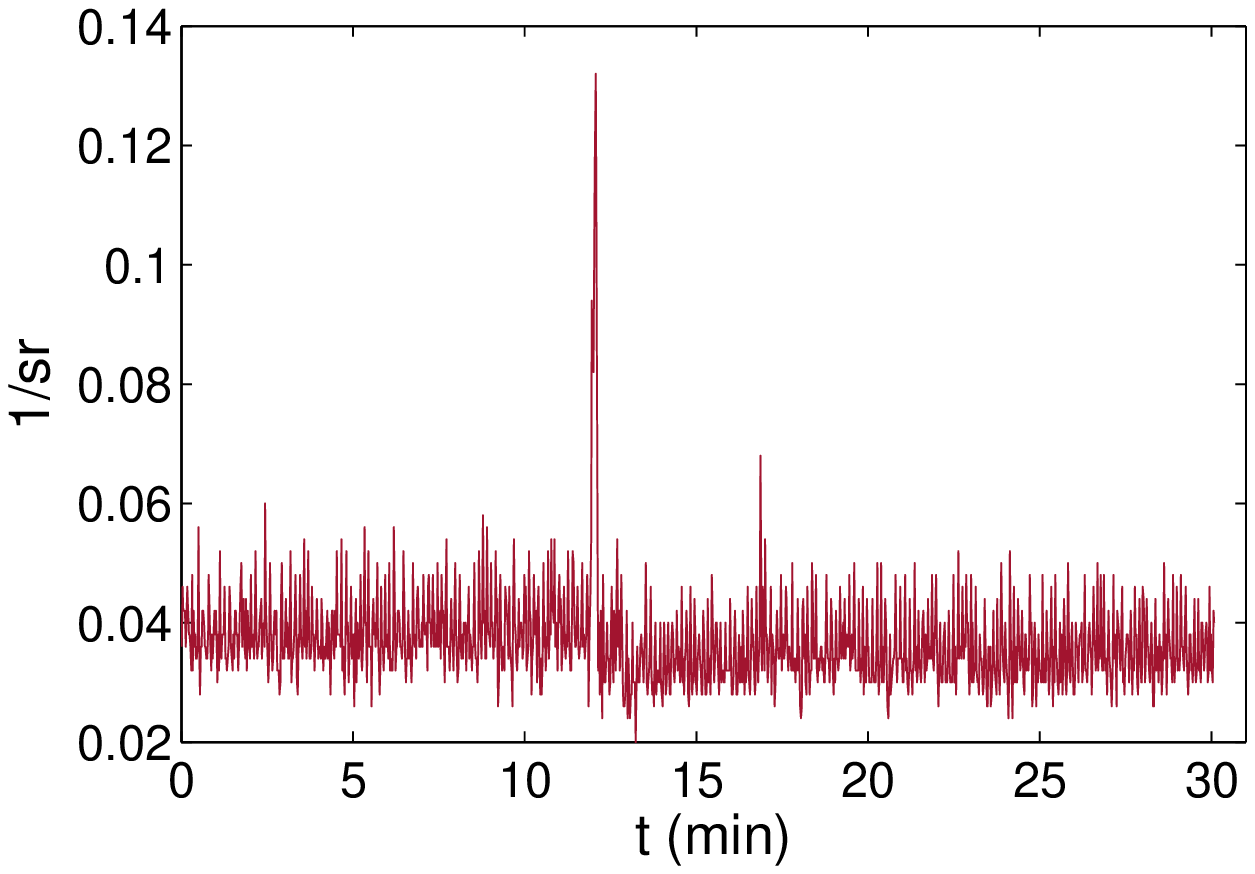}
\includegraphics[width=5cm]{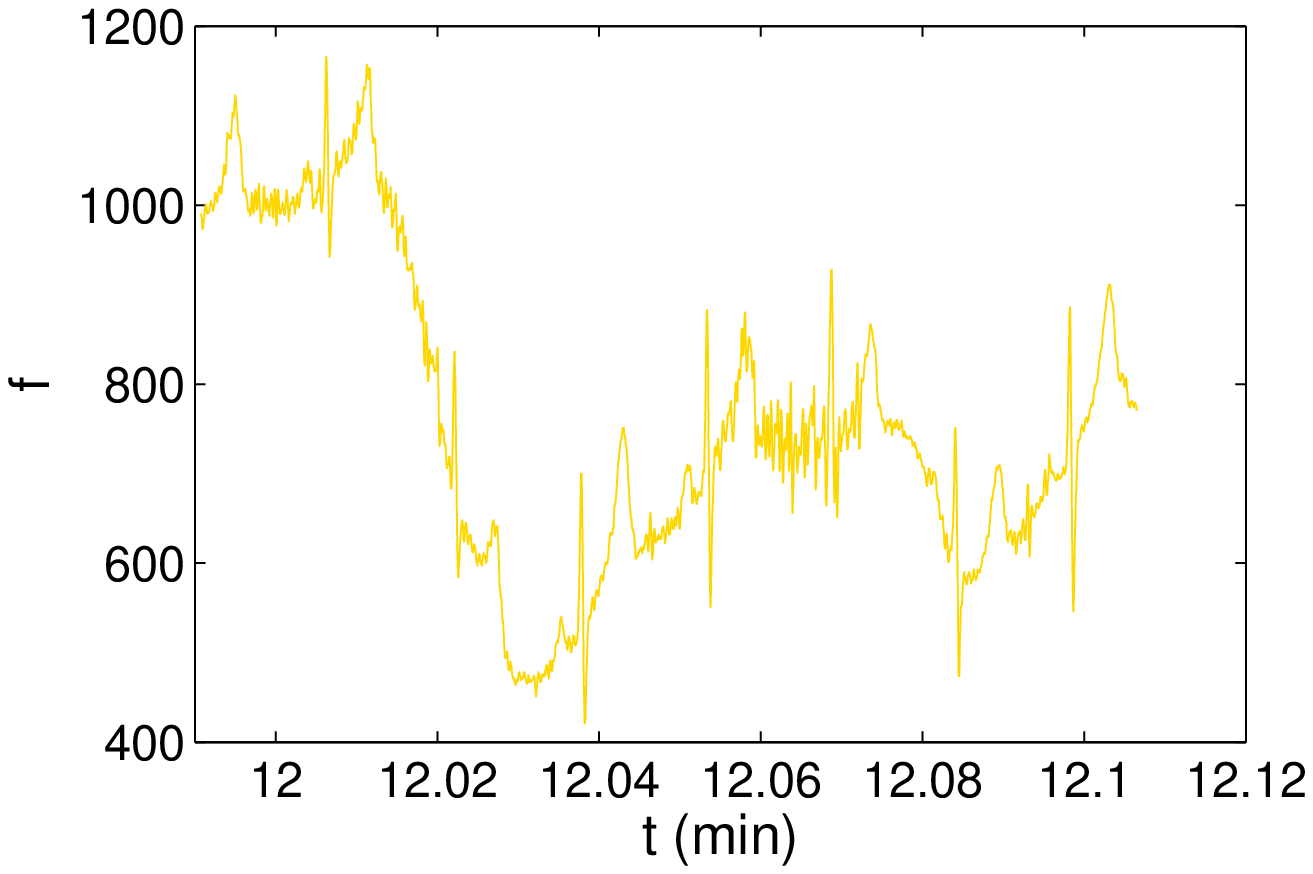}
\includegraphics[width=5cm]{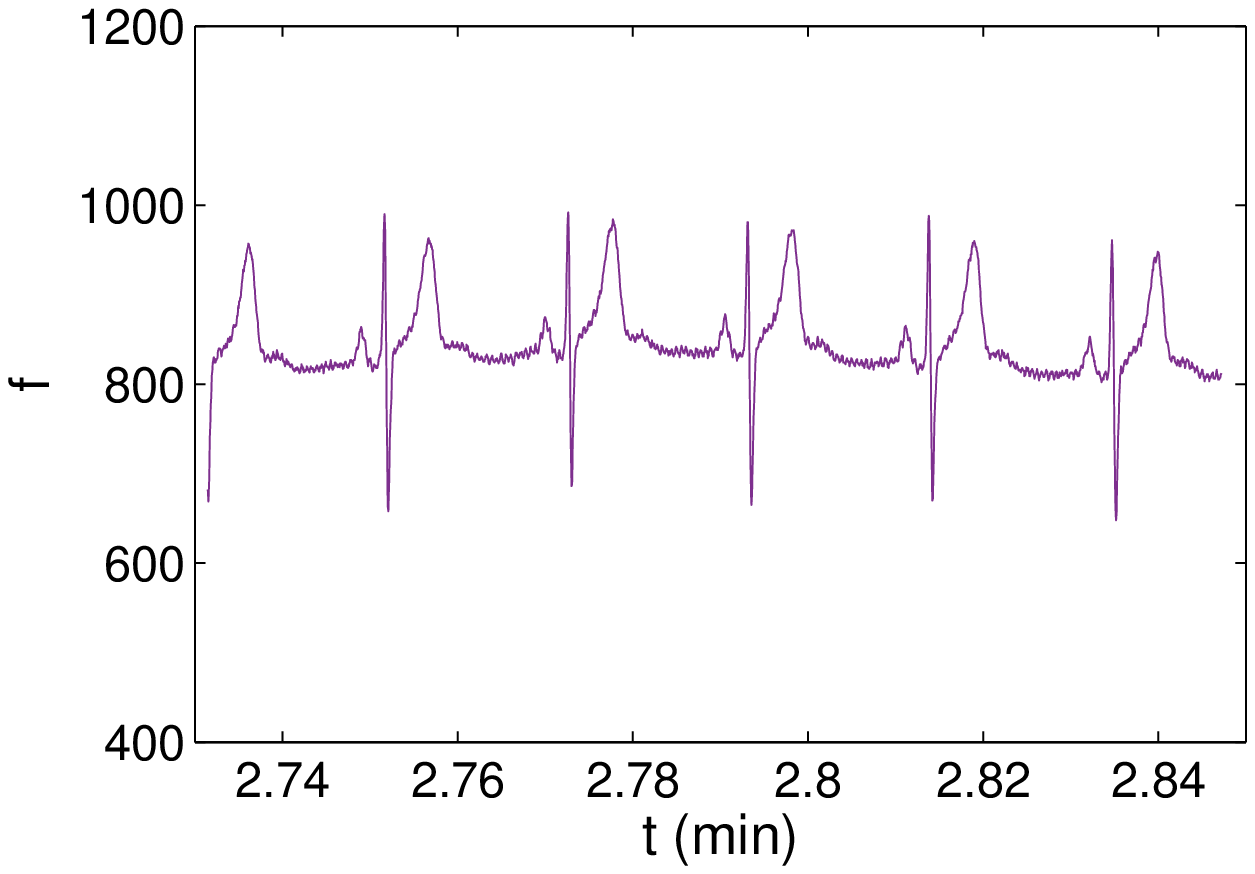}
\caption{The left graph shows the  
 inverse local sparsity values for each segment in the signal 
partition of record 117. The middle graph is the 
piece of signal where the peak of $1/\sr$ takes 
place. The right graph is a piece of signal of the same 
length but at different time interval.}
\label{Fig3}
\end{center}
\end{figure}

From the results shown in Tables~\ref{TABLE4} - 
\ref{TABLE5} we can assert that the dictionary-based approach presented in this paper, which is  powerful for dimensionality reduction at low level distortion, is also relevant to compression. In that sense the 
adopted strategy produces results which significantly
 improve upon the state of the art for the full 
MIT-BIH Arrhythmia data set \cite{LKL11,MZD15,TZW18,PSS16}.  In   
regard to CR the results are close to those recently 
reported in
\cite{LRN19}. In that publication the 
compression results are obtained with an equivalent 
strategy as that described in Sec.~\ref{ES}, but applying 
a FWT on the full signal instead of using a piecewise 
model as proposed here. With respect to 
processing time the approach in \cite{LRN19} is far more 
efficient though. Indeed, with the same equipment 
 the construction of Table~\ref{TABLE2}
 in this paper takes 30 s per record. The equivalent 
table with the approach of \cite{LRN19} takes 0.14 s per 
record.  However, the central aim 
of the present work is to accomplish dimensionality 
reduction. In that sense the dictionary approach is 
much  more effective. As shown in the second column of 
Table~\ref{TABLE1}, to achieve
$\ov{\PRD}=53$ the $\CDF_{\Di}$ method uses 
$84\%$ less components than the {\bf{FWT}} one. 
Moreover, the dictionary approach yields a piecewise 
model which can be useful for analysis. 
It is interesting to note, for instance, that abrupt 
 variations 
of the local sparsity ratio (c.f. \eqref{sr}) gives 
information about the presence of 
 nonstationary noise or significant 
distortions in patters. 
The left graph in Fig.~\ref{Fig3} plots the 
values $1/\sr(q), q=1,\ldots,1300$ for record 117. 
The middle top graph shows the piece of signal 
where the peak of $1/\sr$ takes place and 
the right graph shows a  piece of signal, of 
the same length, outside that time interval.  
\section{Conclusions}
\label{Con}
A method for dimensionality reduction of ECG 
signals has been proposed.
The approach was designed within the framework of
 sparse representation. A number of 
wavelet based dictionaries have been introduced to 
undertake the signal model via the OOMP greedy 
strategy. The resulting representation 
was shown to require significantly 
less elementary component than those needed when using a 
wavelet basis.  The method was proven to 
  be useful for ECG compression at low level distortion.
 The compression results largely surpass  
 benchmarks within the state of the art for 
 the MIT-BIH Arrhythmia database. 

The MATLAB software for constructing the dictionaries and 
reproducing the results in the paper has been made available 
on a dedicated website. Although the numerical construction 
of the signal model is computationally intensive it 
can be effectively realized, even 
in the MATLAB environment, where it typically takes  
30 s to compress a 30 min record and 1 s to 
recover the signal from the compressed file. 
  Moreover, since the 
approximation is carried out independently in every 
segment of the signal partition, there is room for  
  straightforward parallelization using multiprocessors. 
The sensitivity of local sparsity to nonstationary 
noise or significant distortion of patterns leads to conclude that 
the model might be useful to support the development of 
 techniques for analysis of ECG records. 

\section*{Acknowledgments}
 We are indebted  
 to Karl Skretting for making available the
 MATLAB functions
{\tt{Huff06}} 
 which has been used for adding entropy coding to 
the proposed compression scheme.

\end{document}